\documentclass[aps,prb,showpacs,twocolumn,amsmath,amssymb,superscriptaddress,nofootinbib]{revtex4}

\usepackage[english]{babel}
\usepackage{times}
\usepackage{latexsym}
\usepackage{graphics}
\usepackage{subfigure}
\usepackage{epsfig}

\newcommand{\infig}[2]{\begin{center}
                                    \mbox{ \epsfxsize #1 \epsfbox{#2.eps}}
                                      \vspace{-0.8cm}
                                    \end{center}}

\newcommand{\ie}{{\it i.e.}~}
\newcommand{\eg}{{\it e.g.}~}

\newcommand{\kB}{k_\textrm{B}}

\newcommand{\z}{\zeta}

\newcommand{\HBBH}{H_\textrm{BBH}}
\newcommand{\HSBm}{H_\textrm{eff}}

\newcommand{\Ub}{U_\textrm{b}}
\newcommand{\UB}{U_\textrm{B}}
\newcommand{\UbB}{U_\textrm{bB}}

\newcommand{\Jb}{J_\textrm{b}}
\newcommand{\JB}{J_\textrm{B}}

\renewcommand{\b}{\mathbf{b}}
\newcommand{\B}{\mathbf{B}}

\newcommand{\n}{\mathbf{n}}
\newcommand{\N}{\mathbf{N}}

\newcommand{\compB}{\mathbf{\mathcal{B}}}
\newcommand{\compN}{\mathbf{\mathcal{N}}}
\newcommand{\compV}{\mathcal{V}}

\newcommand{\spin}{\mathbf{\sigma}}
\newcommand{\h}{\mathbf{h}}
\newcommand{\e}{\mathbf{e}}

\newcommand{\Sx}{\mathbf{S}^x}
\newcommand{\Sy}{\mathbf{S}^y}
\newcommand{\Sz}{\mathbf{S}^z}

\begin{document}

\title{Disorder versus the Mermin-Wagner-Hohenberg effect:
From classical spin systems \\ to ultracold atomic gases}

\author{J.~Wehr}
\affiliation{ICREA and ICFO-Institut de Ci\`encies Fot\`oniques,
Parc Mediterrani de la Tecnologia,
E-08860 Castelldefels (Barcelona), Spain}
\affiliation{Department of Mathematics, The University of Arizona, Tucson, Arizona 85721-0089, USA}
\author{A.~Niederberger}
\affiliation{ICREA and ICFO-Institut de Ci\`encies Fot\`oniques,
Parc Mediterrani de la Tecnologia,
E-08860 Castelldefels (Barcelona), Spain}
\author{L.~Sanchez-Palencia}
\affiliation{Laboratoire Charles Fabry de l'Institut d'Optique,
CNRS and Univ. Paris-Sud,
campus Polytechnique, RD 128,
F-91127 Palaiseau cedex, France}
\author{M.~Lewenstein}
\affiliation{ICREA and ICFO-Institut de Ci\`encies Fot\`oniques,
Parc Mediterrani de la Tecnologia,
E-08860 Castelldefels (Barcelona), Spain}
\affiliation{Institut f\"ur Theoretische Physik, Universit\"at Hannover, D-30167 Hannover, Germany}

\date{\today}

\begin{abstract}
We propose a general mechanism of {\it random-field-induced order}
(RFIO), in which long-range order is induced by a random field
that breaks the continuous symmetry of the model. We particularly
focus on the case of the classical ferromagnetic $XY$ model on a
2D lattice, in a uniaxial random field. We prove rigorously that
the system has spontaneous magnetization at temperature $T=0$, and
we present strong evidence that this is also the case for small
$T>0$. We discuss generalizations of this mechanism to various
classical and quantum systems. In addition, we propose possible
realizations of the RFIO mechanism, using ultracold atoms in an
optical lattice. Our results shed new light on controversies in
existing literature, and open a way to realize  RFIO with
ultracold atomic systems.
\end{abstract}


\maketitle

\section{Introduction}
\label{sec:introduction}
\subsection{Disordered ultracold quantum gases}
Studies  of disordered systems  constitute a new,
rapidly developing, branch of the physics of ultracold gases.
In condensed matter physics (CM), the role of {\it quenched} (\ie independent of time)
disorder cannot be overestimated: it is present in nearly all CM systems,
and leads to numerous phenomena that dramatically change both qualitative and quantitave
behaviours of these systems. This leads, for instance, to novel thermodynamical and
quantum phases \cite{lifshits1988,akkermans2004},
and to strong phenomena, such as Anderson localization
\cite{anderson1958,localization1d,gang4,vantiggelen1999}.
In general, disorder can hardly be controlled in CM systems. In contrast,
it has been proposed recently, that quenched disorder (or pseudo-disorder) can be
introduced in a controlled way in ultracold atomic systems,
using optical potentials generated by speckle radiations \cite{bodzio,roth,grynberg},
impurity atoms serving as random scatterers \cite{castin},
or quasi-cristalline lattices \cite{laurent-luis}.
This opens fantastic possibilities to investigate the effect of disorder in
controlled systems
(for a review in the context of cold gases, see Ref.~\onlinecite{ahufinger2005}).
Recently, several groups have initiated the experimental study of disorder with
Bose-Einstein condensates (BEC) \cite{inguscio,clement,schulte2005},
and strongly correlated Bose gases \cite{fallani2006}.
In the center of interest of these works is one of the  most fundamental issues
of disordered systems that concerns the interplay
between Anderson localization and interactions in many body Fermi or Bose systems
at low temperatures.
In non-interacting atomic sytems, localization is feasible experimentally \cite{cord},
but even weak interactions can drastically change the scenario.
While weak repulsive interactions tend to delocalize, strong ones in confined geometries
lead to Wigner-Mott-like localization \cite{fisher1989}.
Both experiments and theory indicate that in gaseous systems with large interactions,
stronger localization effects occur in the excitations of a BEC
\cite{clement,pavloff,paul,laurent2006}, rather than on the BEC wavefunction itself.
In the limit of weak interactions, a Bose gase enters a {\it Lifshits glass} phase,
in which several  BECs in various localized single atom orbitals
from the low energy tail of the spectrum coexist \cite{laurent2006b}
(for `traces' of the Lifshits glass in the meanfield theory, see Ref.~\onlinecite{schulte2005}).
Finally, note that disorder in Fermi gases, or in Femi-Bose atomic mixtures, should
allow one to realize various fermionic disorderded phases,
such as a Fermi glass, a Mott-Wigner glass, `dirty' superconductors, etc.
(Ref.~\onlinecite{ahufinger2005}), or even quantum spin glasses \cite{spinglass2004}.

\subsection{Large effects by small disorder}
One of the most appealing effects of disorder is that even
extremely small randomness can have dramatic consequences. The
paradigm example in classical physics is the Ising model for
which an arbitrarily small random magnetic field destroys
magnetization even at temperature T=0 in two dimensions, 2D (Refs.~\onlinecite{imry,aizenman}),
but not in $D>2$ (Ref.~\onlinecite{imbrie}). This result has been generalized to systems with
continuous symmetry in random fields distributed in accordance with this symmetry
\cite{imry,aizenman}.
For instance, the Heisenberg model in a $SO(3)$-symmetrically distributed field
does not magnetize up to 4D.

In quantum physics, the paradigm example of large effects induced by small disorder is
provided by the above-mentioned Anderson
localization which occurs in 1D and 2D in arbitrarily small random
potentials \cite{gang4}. In this paper, we propose an even more
intriguing opposite effect,  where disorder counter-intuitively
favors ordering:
 a general mechanism of {\it random-field-induced order}
(RFIO) by which certain spin models magnetize at a higher
temperature in the presence of arbitrarily small disorder than in
its absence, provided that the disorder breaks the continuous
symmetry of the system.

\subsection{Main results and plan of the paper}
As is well known, as a consequence of the Mermin-Wagner-Hohenberg
theorem \cite{Mermin-Wagner}, spin or field theoretic systems with
continuous symmetry in dimensions less or equal to 2D cannot
exhibit long range order. The mechanism that we propose here
breaks the continuous symmetry, and in this sense acts against the
Mermin-Wagner-Hohenberg no-go rule in 2D. In particular, we prove
rigorously that the classical $XY$ spin model on a 2D lattice in a
uniaxial random field magnetizes spontaneously at $T=0$ in the
{\it direction perpendicular} to the magnetic field axis, and
provide strong evidence that this is also the case at small
positive temperatures. We discuss generalizations of this
mechanism to classical and quantum XY and Heisenberg models in 2D
and 3D. In 3D, the considered systems do exhibit long range order
at finite $T>0$, but in this case the critical temperature
decreases with the `size' of the symmetry group: it is the largest
for the Ising model (the discrete group $Z_2$), higher for the
$XY$ model [the continuous group $U(1)$], and the highest for the
Heisenberg model [the continuous group $SU(2)$, or $SO(3)$]. In
this case we expect that our mechanism will lead to an increase of
the critical temperature for the $XY$ and Heisenberg models, and
to an increase of the order parameter value at a fixed temperature
for the disordered system in comparison to the non-disordered one.
Finally, we propose three possible and experimentally feasible
realizations of the RFIO phenomenon using ultracold atoms in
optical lattices.

The paper is organized as follows.
In section~\ref{sec:rfio}, we present the results concerning the RFIO
in the classical $XY$ model on a 2D lattice.
First, we rigorously prove that the system magnetizes in the direction perpendicular
to the direction of the random magnetic field at $T=0$, and then, we present arguments
that the magnetization persists in small $T>0$ case,
as well as the results of numerical classical Monte Carlo simulations.
Section~\ref{sec:generalizations} is fully devoted to the discussion of
the generalizations of the RFIO mechanism to several other classical and quantum
spin systems.
In section~\ref{sec:ultracold}, we discuss several experimentally feasible
realizations of RFIO in ultracold atomic systems.
Finally, we summarize our results in section~\ref{sec:conclusion}.

\section{RFIO in classical $XY$ model}
\label{sec:rfio}
\subsection{The system under study}
\begin{figure}[t!]
\infig{27em}{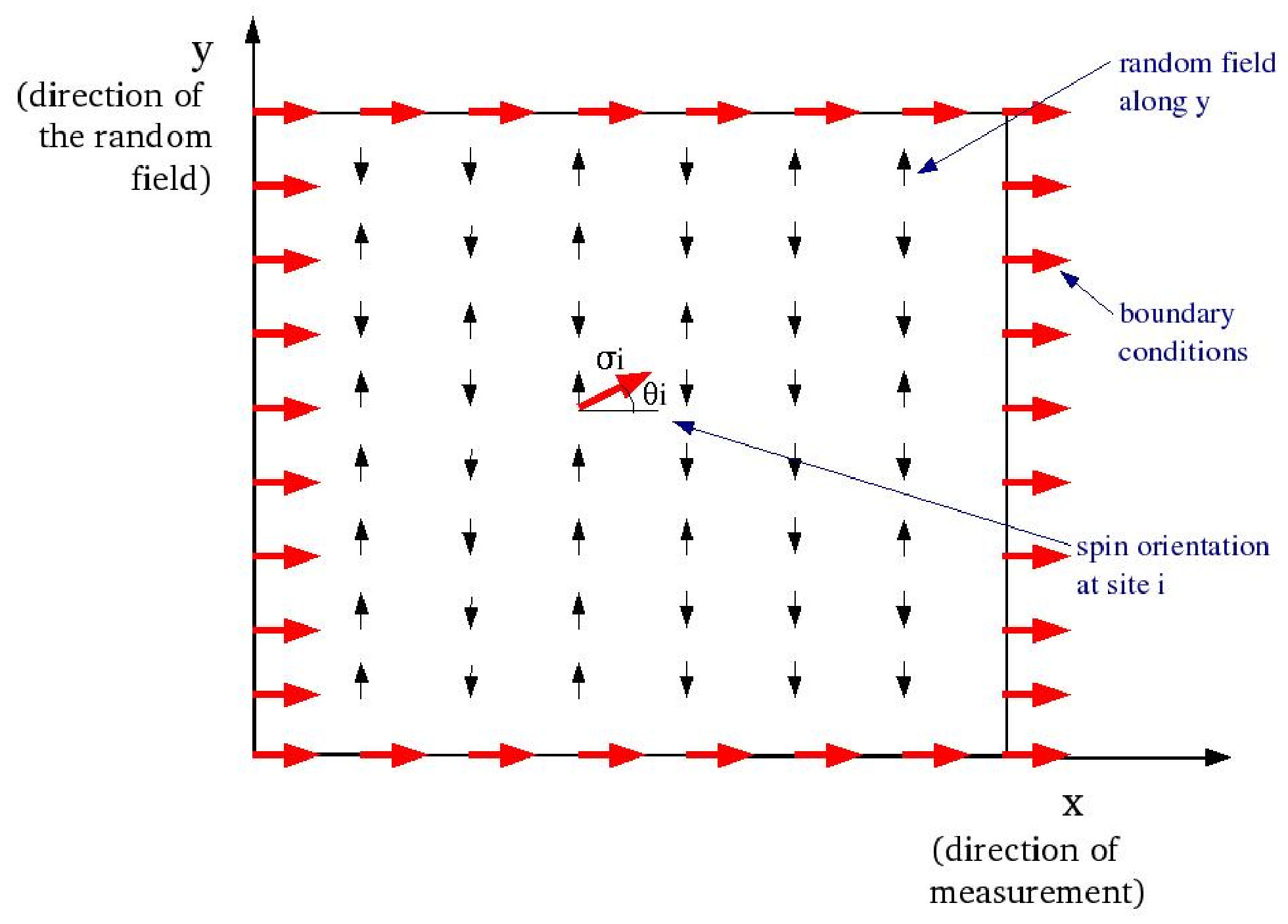}
\caption{(color online) XY model on a 2D square lattice in a random magnetic field.
The magnetic field is oriented along the $y$ axis,
$\h_i = \eta_i {\bf e}_y$, where $\eta_i$ is a real
random number.
Right boundary conditions are assumed on the outer square,
possibly placed at infinity (see text).}
\label{fig:configuration}
\end{figure}

Consider a classical spin system on the 2D square lattice
${\bf Z}^2$, in a random magnetic field, $\h$ (see Fig.~\ref{fig:configuration}).
The spin variable, $\spin_i=(\cos\theta_i,\sin\theta_i)$, at
a site $i \in {\bf Z}^2$ is a unit vector in the $xy$ plane. The
Hamiltonian (in units of the exchange energy $J$) is given by
\begin{equation}
H/J = - \sum_{|i-j| = 1}\spin_i \cdot \spin_j - \epsilon \sum_i
\h_i \cdot \spin_i.
\label{xy}
\end{equation}
Here the first term is the standard nearest-neighbor interaction
of the XY-model, and the second term represents a small random
field perturbation. The $\h_i$'s are assumed to be independent,
identically distributed random, 2D vectors.

For $\epsilon = 0$, the model has no
spontaneous magnetization, $m$, at any positive $T$.  This was first
pointed out in Ref.~\onlinecite{Herring-Kittel}, and later developed
into a class of results known as the Mermin-Wagner-Hohenberg theorem
\cite{Mermin-Wagner} for
various classical, as well as quantum two-dimensional spin systems
with continuous symmetry.  In higher dimensions the system does
magnetize at low temperatures. This follows from the spin wave
analysis \cite{spinwave}, and has been given a rigorous proof in
Ref.~\onlinecite{Frohlich}.  The impact of a random field term
on the behavior of the model
was first addressed in Refs.~\onlinecite{imry,aizenman}, where it was
argued that if the distribution of the random variables $\h_i$ is
invariant under rotations, there is no spontaneous magnetization
at any positive $T$ in any dimension $D \leq 4$.  A rigorous proof
of this statement was given in Ref.~\onlinecite{aizenman}.  Both works use
crucially the rotational invariance of the distribution of the
random field variables.

Here we consider the case where  $\h_i$ is directed along the
$y$-axis: $\h_i = \eta_i {\bf e}_y$, where $\e_y$ is the unit
vector in the $y$ direction, and $\eta_i$ is a random real
number.
Such a random field obviously breaks the continuous symmetry of
the interaction and a question arises whether the model still has
no spontaneous magnetization in two dimensions. This question has
been given contradictory answers in Refs.~\onlinecite{Dotsenko,Minchau}:
while Ref.~\onlinecite{Dotsenko} predicts that a small random field in
the $y$-direction does not change the behavior of the model, 
Ref.~\onlinecite{Minchau} argues that it leads to the presence of spontaneous
magnetization, $m$, in the direction perpendicular to the random field axis in low
(but not arbitrarily low) temperatures. Both works use
renormalization group analysis, with Ref.~\onlinecite{Minchau} starting
from a version of the Imry-Ma scaling argument to prove that the
model magnetizes at zero temperature.

The same model was subsequently studied by Feldman \cite{Feldman},
using ideas similar to the argument given in the present paper. As
we argue below, however, his argument contains an essential gap,
which is filled in the present work.
We first present  a complete proof that the system indeed
magnetizes at $T=0$, and argue that the ground state
magnetization
is stable under inclusion of small thermal fluctuations.
For this, we use a version of the Peierls contour argument
\cite{Peierls}, eliminating first the possibility that Bloch walls
or vortex configurations destroy the transition.

\subsection{RFIO at $T=0$}
Let us start by a rigorous analysis of the ground state. Consider
the system in a square $\Lambda$ with the `right' boundary
conditions, $\spin_i = (1,0)$, for the sites $i$ on the outer
boundary of $\Lambda$ (see Fig.~\ref{fig:configuration}).
The energy of any spin configuration
decreases if we replace the $x$ components of the spins by their
absolute values and leave the $y$ components unchanged. It follows
that in the ground state, $x$ components of all the spins are
nonnegative.
As the size of the system increases, we expect the $x$ component
of the ground state spins to decrease, since they feel less
influence of the boundary conditions and the ground state value of
each spin will converge.  We thus obtain a well-defined
infinite-volume ground state with the `right' boundary
conditions at infinity.

We emphasize that the above convergence statement is nontrivial
and requires a proof.  Physically it is, however, quite natural. A
similar statement has been rigorously proven for ground states of
the Random Field Ising Model using Fortuin-Kasteleyn-Ginibre monotonicity techniques \cite{aizenman,fkg}.

\subsection{Infinite volume limit}
{\it A priori} this infinite-volume ground state could coincide
with the ground state of the Random Field Ising Model, in which
all spins have zero $x$ component. The following argument shows
that this is {\it not} the case. Suppose that the spin $\spin_i$
at a given site $i$ is aligned along the y-axis,
\ie $\cos \theta_i = 0$. Since the derivative of the energy
function with respect to $\theta_i$ vanishes at the minimum, we
obtain
\begin{equation}
\sum_{j:|i-j|=1}\sin(\theta_i - \theta_j) = 0.
\end{equation}
Since $\cos \theta_i = 0$, this implies $ \sum_{j:|i-j|=1}\cos
\theta_j = 0. $ Because in the `right' ground state all spins lie
in the (closed) right half-plane $x \ge 0$, all terms in the above
expression are nonnegative and hence have to vanish. This means
that at all the nearest neighbors $j$ of the site $i$, the ground
state spins are directed along the $y$-axis as well. Repeating
this argument, we conclude that the same holds for all spins of
the infinite lattice, \ie
the ground state is the (unique) Random Field Ising Model ground
state. This, however, leads to a contradiction, since assuming
this, one can construct a field configuration, occurring with a
positive probability, which forces the ground state spins to have
nonzero $x$ components. To achieve this we put strong positive
($\eta_i > 0$) fields on the boundary of a square and strong
negative fields on the boundary of a concentric smaller square. If
the fields are very weak in the area between the two boundaries,
the spins will form a Bloch wall, rotating gradually from $\theta
= {\pi / 2}$ to $\theta = - {\pi / 2}$. Since such a local field
configuration occurs with a positive probability, the ground state
{\it cannot have} zero $x$ components everywhere, contrary to our
assumption.

We would like to emphasize the logical structure of the above
argument, which proceeds indirectly assuming that the ground
state spins (or, equivalently, at least one of them) have zero
$x$ components and reach a contradiction.  The initial
assumption is used in an essential way to argue existence of the
Bloch wall interpolating between spins with $y$ components equal
to $+1$ and $-1$.  It is this part of the argument that we think
is missing in Ref.~\onlinecite{Feldman}.  Note, that this argument applies to
strong, as well as to weak random fields, so that the ground state
is never, strictly speaking, field-dominated and always exhibits
magnetization in the $x$-direction.  Moreover, the argument does
not depend on the dimension of the system, applying in particular
in one dimension. We argue below that in dimensions greater than
one the effect still holds at small positive temperatures, the
critical temperature depending on the strength of the random field
(and presumably going to zero as the strength of the field
increases).

\subsection{RFIO at low positive $T$}
To study the system at low positive $T$, we need to ask what are
the typical low energy excitations from the ground state. For
$\epsilon = 0$, continuous symmetry allows Bloch walls, \ie
configurations in which the spins rotate gradually over a large
region, for instance from left to right.  The total excitation
energy of a Bloch wall in 2D is of order one, and it is the
presence of such walls that underlies the absence of continuous
symmetry breaking.  However, for $\epsilon > 0$, a Bloch wall
carries additional energy, coming from the change of the direction
of the $y$ component of the spin, which
is proportional to the area of the wall (which is of the order
$L^2$ for a wall of linear size $L$ in two dimensions), since the
ground state spins are adapted to the field configuration, and
hence overturning them will increase the energy per site.
Similarly, vortex configurations, which are important low-energy
excitations in the nonrandom XY model, are no longer energetically
favored in the presence of a uniaxial random field.

We are thus left, as possible excitations, with sharp domain
walls, where the $x$ component of the spin changes sign rapidly.
To first approximation we consider excited configurations, in
which spins take either their ground state values, or the
reflections of these values in the $y$-axis.  As in the standard
Peierls argument \cite{Peierls}, in the presence of the right
boundary conditions, such configurations can be described in terms
of contours $\gamma$ (domain walls), separating spins with
positive and negative $x$ components. If $m_i$ is the value of the
$x$ component of the spin $\sigma_i$ in the ground state with the
right boundary conditions, the energy of a domain wall is the sum
of $m_im_j$ over the bonds $(ij)$ crossing the boundary of the
contour.  Since changing the signs of the $x$ components of the
spins does not change the magnetic field contribution to the
energy, the Peierls estimate shows that the probability of such
a contour is bounded above by $\exp(-2\beta\sum_{(ij)}m_im_j), $
with $\beta = {J/ k_BT}$.

We want to show that for a typical realization of the field, $\h$,
(\ie with probability one), these probabilities are summable, \ie
their sum over all contours containing the origin in their
interior is finite. It then follows that at a still lower $T$,
this sum is small, and the Peierls argument proves that the system
magnetizes (in fact, a simple additional argument shows that
summability of the contour probabilities already implies the existence
of spontaneous $m$). To show that a series of random variables is
summable with probability one, it suffices to prove the summability of
the series of the expected values.  We present two arguments for
the last statement to hold.

If the random variables $m_i$ are bounded away from zero, \ie $m_i>\sqrt{c}$, for some $c>0$, the
moment generating function of the random variable
$\sum_{(ij)}m_im_j$ satisfies
\begin{equation}
{\bf E}\Big[ \exp \Big(-\beta\sum_{(ij)}m_im_j \Big) \Big] \leq \exp[-c \beta L(\gamma)],
\label{3}
\end{equation}
with $L(\gamma)$ denoting the length of the contour $\gamma$. The
sum of the probabilities of the contours enclosing the origin is
thus bounded by $ \sum_{\gamma}\exp[-c\beta L(\gamma)]. $ The
standard Peierls-Griffiths bound proves the desired summability.

The above argument does not apply if the distribution of the ground state, $m$, contains
zero in its support.  For unbounded distribution of the random field this may very well
be the case, and then another argument is needed.
If we assume that the terms in the sum $\sum_{(ij)}m_im_j$ are independent and
identically distributed, then
${\bf E}[\exp(-2\beta\sum_{(ij)}m_im_j)] =
{\bf E}[\exp(-2\beta m_im_j)]^{L(\gamma)}
= \exp\{L(\gamma) \log{\bf E}[\exp(-2 \beta m_im_j)]\}$
and we just need to observe that
${\bf E}[\exp(-2 \beta m_im_j)] \to 0$
as
$\beta \to \infty$ (since the expression under the expectation sign goes pointwise to
zero and lies between $0$ and $1$) to conclude that
${\bf E}[\exp(-2\beta\sum_{(ij)}m_im_j)]$ behaves as
$\exp[-g(\beta)L(\gamma)]$ for a positive
function
$g(\beta)$ with
$g(\beta) \to \infty$ as $\beta \to \infty$.  While $m_im_j$
are not, strictly speaking, independent, it is natural to assume that their dependence
is weak, \ie their correlation decays fast with the distance of the corresponding
bonds $(ij)$. The behavior of the moment generating function of their sum is then
qualitatively the same, with a renormalized rate function $g(\beta)$, still diverging as
$\beta \to \infty$.
As before, this is enough to carry out the Peierls-Griffiths
estimate which implies spontaneous magnetization in the
$x$-direction.  We remark that our assumption about the fast decay of
correlations implies that the sums of $m_im_j$ over subsets of
${\bf Z}^2$ satisfy a large deviation principle analogous to that
for sums of independent random variables and the above argument
can be restated using this fact.

\subsection{Numerical Monte Carlo simulations}
Based on the above discussion it is expected that the RFIO effect
predicted here will lead to the appearance of magnetization, $m$,
in the $x$ direction of order 1 at low temperatures in systems
much larger than the correlation length of typical excitations.
For small systems, however, the effect may be obscured by finite
size effects, which, due to long-range power law decay of
correlations, are particularly strong in the $XY$ model in 2D. In
particular, the 2D-XY model shows finite magnetization ($m$) in
small systems \cite{bram}, so that RFIO is expected to result in
an increase of the magnetization.

\begin{figure}[t!]
\infig{22em}{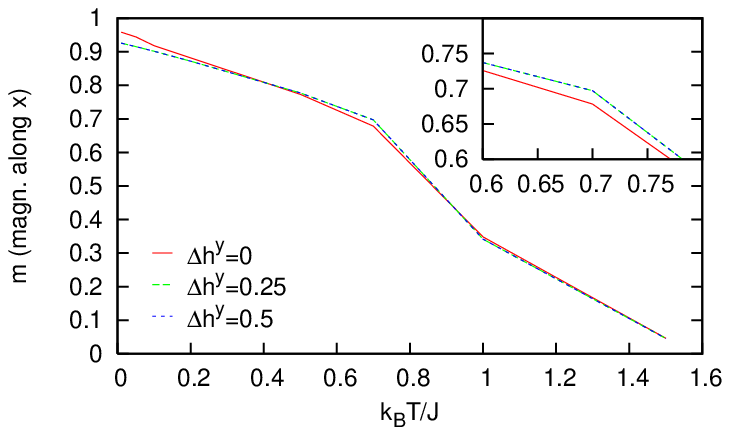}
\caption{(color online)
Results of the Monte-Carlo simulation for the classical 2D-XY model
in a 200$\times$200 lattice.
The Inset is a magnetification of the main figure close to $T=0.7J/\kB$.}
\label{fig:MC}
\end{figure}

We have performed numerical Monte-Carlo simulations \cite{alps}
for the 2D-XY classical model [Hamiltonian~(\ref{xy}), with
$\epsilon = 1$]. We generate a random magnetic field, $\h_i =
\eta_i \e_y$ in the $y$ direction. The $\eta_i$'s are independent
random real numbers, uniformly distributed in $[-\sqrt{3}\Delta
h^y,\sqrt{3}\Delta h^y]$. Note that $\Delta h^y$ is thus the
standard deviation of the random field $\h_i$. Boundary conditions
on the outer square correspond to $\sigma_i = (1,0)$ [see
Fig.~\ref{fig:configuration}]. The calculations were performed in
2D lattices with up to 200$\times$200 lattice sites for various
temperatures. The results are presented in Fig.~\ref{fig:MC}.

At very small temperature, the system magnetizes in the absence of
disorder ($m$ approaches 1 when T tends to 0) due to the finite
size of the lattice \cite{bram}. In this regime, a random field in
the $y$ direction tends to induce a small local magnetization,
parallel to $\h_i$, so that the magnetization in the $x$
direction, $m$, is slightly reduced. At higher temperatures ($T
\simeq 0.7J/\kB$ in Fig.~\ref{fig:MC}), the magnetization is
significantly smaller than 1 in the absence of disorder. This is
due to non-negligible spin wave excitations. In the presence of
small disorder, these excitations are suppressed due to the RFIO
effect discussed in this paper. We indeed find that, at
$T=0.7J/\kB$, $m$ increases by 1.6\% in presence of the uniaxial
disordered magnetic field. At larger temperatures, excitations,
such as Bloch walls or vortices are important and no increase of
the magnetization is found when applying a small random field in
the $y$ direction.

\section{RFIO in other systems}
\label{sec:generalizations} The RFIO effect predicted above may be
generalized to other spin models, in particular those that have
finite correlation length.
 Here we list the most spectacular generalizations:

\subsection{2D Heisenberg ferromagnet (HF) in random fields of
various symmetries}
Here the interaction has the same
form as in the $XY$ case, but spins take values on a unit sphere.
As for the $XY$ Hamiltonian,
if the random field distribution has the
same symmetry as the interaction part,
\ie if it is symmetric under rotations in three dimensions, the
model has no spontaneous magnetization up to 4D
(see Ref.~\onlinecite{imry,aizenman}). If the random field is uniaxial, \eg
oriented along the $z$ axis, the system still has a continuous
symmetry (rotations in the $xy$ plane), and thus cannot have
spontaneous magnetization in this plane.  It cannot magnetize in
the $z$ direction either, by the results of Ref.~\onlinecite{aizenman}.
Curiously enough, a field distribution with an intermediate
symmetry may lead to symmetry breaking. Namely, arguments fully
analogous to the previous ones
imply that if the random field
takes values in the $yz$ plane with a distribution invariant
under rotations,
the system will magnetize in the
$x$ direction.  We are thus faced with the possibility that a planar
field distribution breaks the symmetry, while this is broken neither by
a field with a spherically symmetric distribution nor by a
uniaxial one.

\subsection{3D $XY$ and Heisenberg models in a random field
of various symmetries}
We have argued that the 2D $XY$ model with
a small uniaxial random field orders at low $T$.  Since in the
absence of the random field spontaneous magnetization occurs only
at $T=0$, this can be equivalently stated by saying that a small
uniaxial random field raises the critical temperature $T_c$ of the
system. By analogy, one can expect that the (nonzero) $T_c$  of
the $XY$ model in 3D becomes higher and comparable to that of the
3D Ising model, in the presence of a small uniaxial field.
A simple meanfield estimate suggests that $T_c$ might increase by
a factor of 2. The analogous estimates for the Heisenberg model in
3D suggest an increase of $T_c$ by a factor 3/2 (or 3)  in a small
uniaxial (or planar rotationally symmetric) field respectively.
These conjectures are the subject of a forthcoming project.

\subsection{Antiferromagnetic systems}
By flipping every
second spin, the classical ferromagnetic models are equivalent to
antiferromagnetic ones (on bipartite lattices). This equivalence
persists in the presence of a random field with a distribution
symmetric with respect to the origin. Thus the above discussion of
the impact of random fields on continuous symmetry breaking in
classical ferromagnetic models translates case by case to the
antiferromagnetic case.

\subsection{Quantum systems}
All of the effects predicted
above should, in principle have quantum analogs. Quantum
fluctuations might, however, destroy the long-range
order, so each of the discussed models should be carefully
reconsidered in the quantum case. Some models, such as the quantum
spin $S=1/2$ Heisenberg model, for instance,  have been widely
studied in literature \cite{anti}.  The Mermin-Wagner theorem
\cite{Mermin-Wagner} implies that the model has no spontaneous
magnetization at positive temperatures in 2D. For $D>2$ spin wave
analysis \cite{bloch1930,dyson1956,ashcroft1976} 
shows the existence of spontaneous magnetization (though a
rigorous mathematical proof of this fact is still lacking).  In
general, one does not expect major differences between the
behaviors of the two models at $T \neq 0$.
It thus seems plausible that the presence of a random field in the
quantum case is going to have effects similar to those in the
classical Heisenberg model. Similarly, one can consider the
quantum Heisenberg antiferromagnet (HAF) and expect phenomena
analogous to the classical case, despite the fact that  unlike
their classical counterparts, the quantum HF and HAF systems are
no longer equivalent. We expect to observe spontaneous staggered
magnetization  in a random uniaxial XY model, or random planar
field HF. A possibility that a random field in the $z$-direction
can enhance the antiferromagnetic order in the $xy$ plane has been
pointed out in Ref.~\onlinecite{huscroft1997}.

\section{Towards the experimental realization of RFIO in ultracold atomic systems}
\label{sec:ultracold} Further understanding of the phenomena
described in this paper will benefit from experimental
realizations and investigations of the above-mentioned models.
Below, we discuss the possibilities to design quantum simulators for
these quantum spin systems using ultracold atoms in optical
lattices (OL).

\subsection{Two-component lattice Bose gas}
\begin{figure}[t!]
\infig{20em}{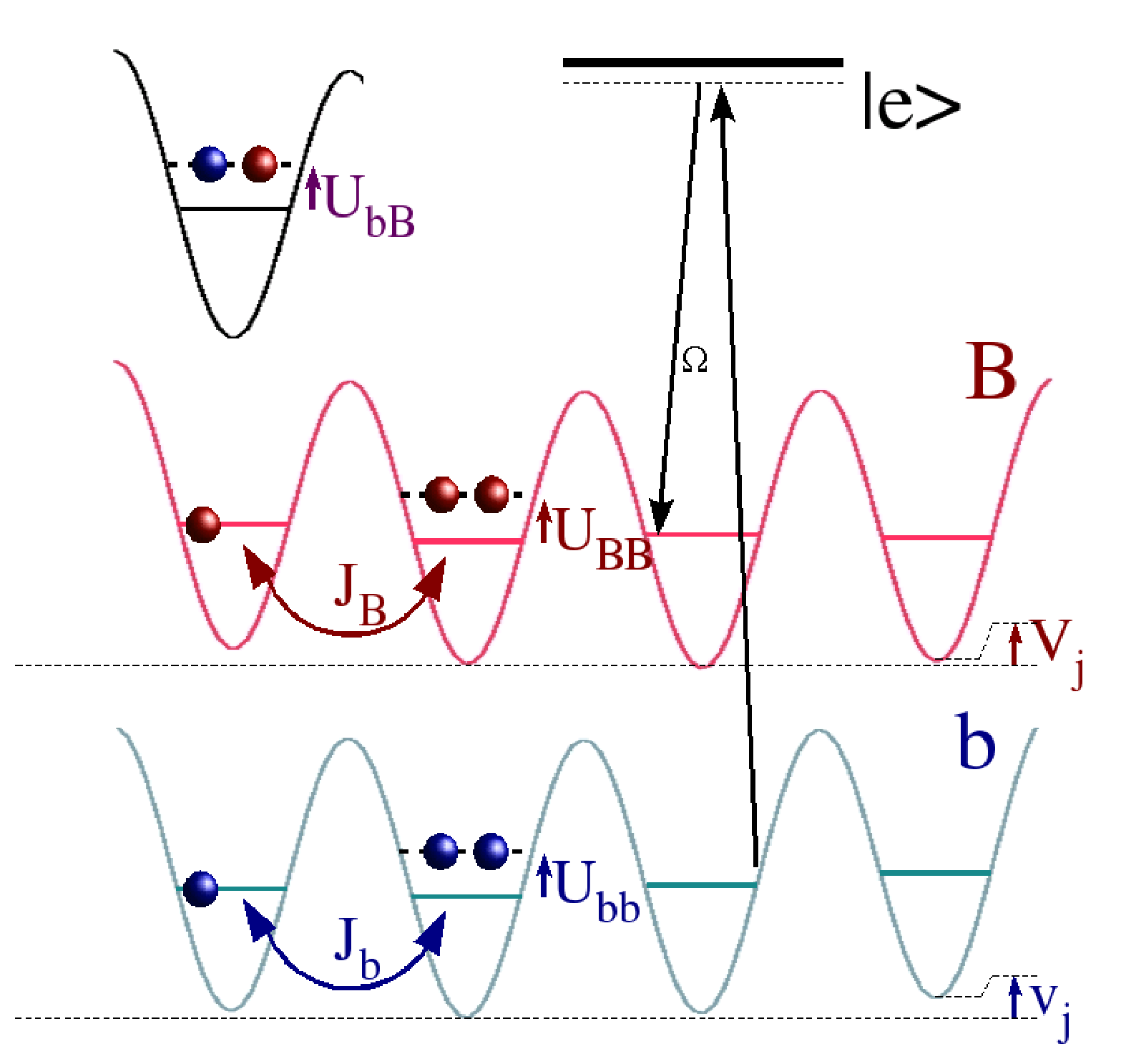}
\caption{(color online) Atomic level scheme of a two-component Bose mixture in a random
optical lattice used to design spin models in random magnetic fields (see text).}
\label{fig:system}
\end{figure}

Consider a two-component  Bose gas confined in an OL with on-site
inhomogeneities. The two components correspond here to two
internal states of the same atom. The low-T physics is captured by
the Bose-Bose Hubbard model (BBH) \cite{jaksch1998} (for a review
of ultracold lattice gases see Ref.~\onlinecite{lew-review}):
\begin{eqnarray}
\HBBH =
  \sum_{j}\Big[ \frac{\Ub}{2} \n_j (\n_j-1)
                +\frac{\UB}{2} \N_j (\N_j-1) \nonumber \\
                +\UbB \n_j \N_j \Big]
 +\sum_{j}\left(v_j \n_j + V_j \N_j \right) \label{eq:BBHm} \\
 -\sum_{\langle j,l \rangle}
   \left[\left(\Jb \b^{\dagger}_j \b^{}_l
         + \JB \B^{ \dagger}_j \B^{}_l\right) + \textrm{h.c.}\right] \nonumber \\
 - \sum_{j}\left(\frac{\Omega_j}{2} \b_j^\dagger \B_j + \textrm{h.c.}\right) \nonumber
\end{eqnarray}
where $\b_j$ and $\B_j$ are the annihilation operators for both
types of Bosons in the lattice site $j$, $\n_j =\b^{\dagger}_j
\b_j$ and $\N_j =\B^{\dagger}_j \B_j$ are the corresponding number
operators, and $\langle j,l \rangle$ denote a pair of adjacent
sites in the OL. In Hamiltonian~(\ref{eq:BBHm}), (i) the first
term describes on-site interactions, including interaction between
Bosons of different types, with energies $\Ub$, $\UB$ and $\UbB$;
(ii) the second accounts for on-site energies; (iii) the third
describes quantum tunneling between adjacent sites and (iv) the
fourth transforms one Boson type into the other with a probability
amplitude $|\Omega|/\hbar$. The last term can be implemented with
an optical two-photon Raman process if the two Bosonic `species'
correspond to two internal states of the same atom (see also
Fig.~\ref{fig:system}). Possibly, both on-site energies $v_j$,
$V_j$ and the Raman complex amplitude $\Omega_j$ can be made
site-dependent using speckle laser light
\cite{inguscio,clement,schulte2005}.

Consider the limit of strong repulsive interactions
($0 < \Jb,\JB,|\Omega_j| \ll U_\textrm{b},U_\textrm{B},U_\textrm{Bb}$)
and a total filling factor of $1$
(\ie the total number of particles equals the number of lattice sites).
Proceeding as in the case of Fermi-Bose mixtures, recently analyzed
by two of the authors in Ref.~\onlinecite{spinglass2004}, we derive an effective
Hamiltonian, $\HSBm$, for the Bose-Bose mixture. In brief, we restrict
the Hilbert space to a subspace $\mathcal{E}_0$ generated by
$\{\prod_j|n_j,N_j\rangle\}$ with $n_j+N_j=1$ at each lattice site,
and we incorporate the tunneling terms via second-order
perturbation theory as in Ref.~\onlinecite{spinglass2004}. We then end up with
\begin{eqnarray}
\HSBm & = &
-\sum_{\langle j,l \rangle}
     {\left(J_{j,l} \compB_j^\dagger \compB_l
      + \textrm{h.c} \right)}
+ \sum_{\langle j,l \rangle}
       {K_{j,l} \compN_j \compN_l} \nonumber \\
& &
+ \sum_{j}{\compV_j \compN_j}
- \sum_{j}{\left(\frac{\Omega_j}{2}\compB_j+ \textrm{h.c.}\right)}
\label{eq:effSBHm}
\end{eqnarray}
where $\compB_j = \b^{\dagger}_j \B_j \mathcal{P}$, $\mathcal{P}$
is the projector onto $\mathcal{E}_0$ and
$\compN_j=\compB^\dagger_j\compB_j$. Hamiltonian $\HSBm$ contains
(i) a hopping term, $J_{j,l}$, (ii) an interaction term between
neighbour sites, $K_{j,l}$, (iii) inhomogeneities, $\compV_j$, and
(iv) a creation/annihilation term. Note that the total number of
composites is not conserved except for a vanishing $\Omega$. The
coupling parameters in Hamiltonian~(\ref{eq:effSBHm}) are
\footnote{The coupling parameters are the same as calculated in
Refs.~\onlinecite{spinglass2004,ahufinger2005} except for the third term
in Eq.~(\ref{eq:effSBHmbis2}) which corresponds to a virtual state
with two $B$ bosons in the same lattice site---forbidden for
Fermions.}:
\begin{eqnarray}
J_{j,l} & = & \frac{J_b J_B}{U_{bB}}
\left[
  \frac{1}{1-\left(\frac{\delta_{j,l}}{U_{bB}}\right)^2}
 +\frac{1}{1-\left(\frac{\Delta_{j,l}}{U_{bB}}\right)^2} \right]
\label{eq:effSBHmbis1} \\
K_{j,l} & = & -\frac{4 J_b^2/U_b}{1-\left(\frac{\delta_{j,l}}{U_b}\right)^2}
              +\frac{2 J_b^2/U_{bB}}{1-\left(\frac{\delta_{j,l}}{U_{bB}}\right)^2} \nonumber \\
& &           -\frac{4 J_B^2/U_B}{1-\left(\frac{\Delta_{j,l}}{U_B}\right)^2}
              +\frac{2 J_B^2/U_{bB}}{1-\left(\frac{\Delta_{j,l}}{U_{bB}}\right)^2} \label{eq:effSBHmbis2} \\
\compV_j & = & V_j-v_j
             + \sum_{\langle j,l \rangle}\left[
           \frac{4 J_b^2/U_b}{1-\left(\frac{\delta_{j,l}}{U_b}\right)^2}
              -\frac{J_b^2/U_{bB}}{1-\frac{\delta_{j,l}}{U_{bB}}}\right. \nonumber \\
& &           \hspace{1.8cm} \left. -\frac{J_B^2/U_{bB}}{1+\frac{\Delta_{j,l}}{U_{bB}}}
              +\frac{4 J_B^2/U_{B}}{1-\left(\frac{\Delta_{j,l}}{U_{B}}\right)^2}
         \right] \label{eq:effSBHmbis3}
~,
\end{eqnarray}
where $\delta_{j,l}=v_j-v_l$ and $\Delta_{j,l}=V_j-V_l$.
Hamiltonian $\HSBm$ describes the dynamics of {\it composite particles} whose
annihilation operator at site $j$ is
$\compB_j=\b^{\dagger}_j \B_j \mathcal{P}$.
In contrast to the  case of Fermi-Bose mixtures discussed in Ref.~\onlinecite{spinglass2004},
where the composites are fermions, in the present case of Bose-Bose mixtures,
they are {\it composite Schwinger Bosons} made of one $B$ boson and one $b$ hole.

Since the commutation relations of $\compB_j$ and
$\compB^\dagger_j$ are those of Schwinger Bosons
\cite{auerbach1994}, we can directly turn to the spin
representation \cite{auerbach1994} by defining
$\Sx_j+i\Sy_j=\compB_j$ and $\Sz_j=1/2-\compN_j$, where
$\compN_j=\compB^\dagger_j\compB_j$. It is important to note that
since Raman processes can convert $b$ Bosons into $B$ Bosons (and
conversely), $\sum_j \langle \compN_j \rangle$ {\it is not fixed}
by the total number of Bosons of each species, \ie the $z$
component of $m$, $\sum_j \langle \Sz_j \rangle$ is not
constrained. For small inhomogeneities
($\delta_{j,l}=v_j-v_l,\Delta_{j,l}=V_j-V_l \ll \Ub, \Ub, \UbB$),
Hamiltonian $\HSBm$ is then equivalent to the anisotropic
Heisenberg $XXZ$ model \cite{auerbach1994} in a random field:
\begin{eqnarray}
H_\textrm{eff} & = &
-J_\bot \sum_{\langle j,l \rangle}
             {\left(\Sx_j \Sx_l + \Sy_j \Sy_l \right)}
- J_z   \sum_{\langle j,l \rangle}
             {\Sz_j \Sz_l} \nonumber \\
& & - \sum_j{\left( h_j^x \Sx_j + h_j^y \Sy_j + h_j^z \Sz_j \right)}
\label{eq:XXZHm}
\end{eqnarray}
where
\begin{eqnarray}
& & J_\bot = \frac{4 \Jb \JB}{\UbB} \label{eq:effSBHmter1} \\
& & J_z = 2 \left( \frac{2\Jb^2}{\Ub} + \frac{2\JB^2}{\UB} - \frac{\Jb^2+\JB^2}{\UbB} \right) \label{eq:effSBHmter2} \\
& & h_j^x = \Omega_j^\textrm{R} \textrm{~~~;~~~}
    h_j^y = -\Omega_j^\textrm{I} \textrm{~~~;~~~}
    h_j^z = \compV_j - \z J_z/2 ~,
\end{eqnarray}
with $\z$
the lattice coordination number,
$\compV_j = V_j-v_j + \z [ 4\Jb^2/\Ub + 4\JB^2/\UB - (\Jb^2+\JB^2)/\UbB ]$
and
$\Omega_j=\Omega_j^\textrm{R}+i\Omega_j^\textrm{I}$.
In atomic systems, all these (possibly site-dependent)
terms can be controlled almost at will
\cite{spinglass2004,lew-review,hubbard}. In particular, by employing various
possible control tools one can reach the
HF ($J_\bot=J_z$) and $XY$ ($J_z=0$) cases.

\subsection{Bose lattice gas embedded in a BEC}
The quantum ferromagnetic $XY$ model in a random field may be
alternatively obtained using the same BBH model, but with strong
state dependence of the optical dipole forces. One can imagine a
situation in which one-component (say $b$) is in the strong
interaction limit, so that only one $b$ atom at a site is
possible, whereas the other ($B$) component is in the Bose
condensed state and provides only a coherent BEC `background' for
the $b$-atoms. Mathematically speaking, this situation is
described by Eq. (\ref{eq:BBHm}), in which $n_i$'s can be equal to
0 or 1 only, whereas $B_i$'s
can be replaced by a classical complex
field (condensate wave function). In this
limit the spin $S=1/2$ states can be associated with the presence, or absence of a $b$-atom in a given site.
In this way, setting $v_j=0$ and $\Omega^{\rm I}_j=0$, one obtains the quantum version of the
$XY$ model (\ref{xy}) with $J=J_{\rm b}$ and a uniaxial random field in the $x$ direction with the strength
determined by $\Omega_j^{\rm R}$.

\subsection{Two-component Fermi lattice gas}
Finally, the $S=1/2$  antiferromagnetic Heisenberg model may be
realized with a Fermi-Fermi mixture at half filling for each
component. This implementation might be of special importance for
future experiments with Lithium atoms. As recently calculated
\cite{werner}, the critical temperature for the N\'eel state in a
3D cubic lattice is of the order of 30nK. It is well known that in
a 3D cubic lattice the critical temperatures for the
antiferromagnetic Heisenberg, the $XY$ and the Ising models are
$T_{c,XY}\simeq 1.5T_{c,Heis}$, and $T_{c,Ising}\simeq
2T_{c,Heis}$. The estimates of these critical temperatures can be,
for instance, obtained applying  the Curie-Weiss mean field method
to the classical models. Suppose that we put the Heisenberg
antiferromagnet in a uniaxial (respectively, planar) random field,
created using the same methods as discussed above, \ie we break
the $SU(2)$ symmetry and put the system into the universality
class of $XY$ (respectively, Ising) models. Mean field estimates
suggest then that we should expect the increase of the critical
temperature by factor 1.5 (respectively, 2), that is up to
$\simeq$ 45 (respectively, 90)nK. Even if these estimates are too
optimistic, and the effect is two, three times smaller, one should
stress, that even an increase by, say 10nK, is of great
experimental relevance and could be decisive for achieving of
antiferromagnetic state.

We would like to stress that similar proposals, as the three discussed above,  have been formulated before \cite{spinprop},
but none of them
treat simultaneously essential aspects for the present schemes:
i) disordered fields, but not bonds;
ii)  arbitrary directions of the fields;
iii) possibility of exploring Ising, $XY$ or Heisenberg symmetries;
iv) realizing the coherent source of atoms;
and
v) avoiding constraints on the magnetization along the $z$ axis.

It is also worth commenting on what are the most important
experimental challenges that have to be addressed in order to
achieve RFIO. Evidently, for the  proposals involving the strong
interaction limit of two-component Bose, or Fermi systems, the
main issue is the temperature which has to be of order of tens of
nano-Kelvins. Such temperatures are starting to be achievable
nowadays (for a carefull discussion in the context of Fermi-Bose
mixtures see Ref.~\onlinecite{henning}), and there exist several proposals
for supplementary cooling of lattice gases, using laser (photons)
or couplings to ultracold BEC (phonon cooling) that can help (for
reviews see Ref.~\onlinecite{lew-review,hubbard}.

\section{Summary}
\label{sec:conclusion}
In this paper, we have proposed  a general mechanism of {\it random-field-induced order}
(RFIO), occurring in systems with continuous symmetry, placed in a random field that breaks,
or reduces this symmetry. We have presented rigorous results
for the case of the 2D-classical ferromagnetic $XY$ model
in a random uniaxial field, and proved that the system has spontaneous
magnetization at temperature $T=0$. We have  presented also a rather strong evidence
that this is also the case for small $T>0$. Several generalizations of this mechanism
to various classical and quantum systems were discussed.
We have presented also detailed proposals to realize RFIO in experiments using
two-component Bose lattice gases, one-component Bose lattice gases embedded in BECs, or
two-component Fermi lattice gases. Our results shed light on controversies in existing
literature, and open the way to realize  RFIO with ultracold atoms
in an optical lattice.

It is worth mentioning two further realizations of RFIO studied by
us recently. RFIO occurs in a two-component trapped Bose gas at
$T=0$, when the gas is condensed and the two components are
coupled by Raman transition of random strength, but fixed phase.
Although such a system belongs to the universality class of the
(trapped, \ie located in an inhomogenous field) $XY$ model, it
exhibits the RFIO effect in a  much stronger manner than the $XY$
model discussed in the present paper. We have found this
observation important enough to devote a separate detailed paper
to it \cite{sacha}. Similarly, we have studied numerically RFIO in
1D for quantum $XY$ and Heisenberg chains \cite{dziarmaga2006}. In such systems, even
at $T=0$, magnetization vanishes, but amazingly enough the RFIO
effect seems to work at the level of the magnetic
susceptibilities. Adding a random field confined to a certain axis
(respectively, plane), {\it increases} siginificantly the magnetic
suceptibility in the perpedicular directions (respectively,
direction).

\acknowledgements
We thank I. Bloch, T. Roscilde, and C. Salomon for very enlightening discussions.
We acknowledge  the support of DFG, ESF Programme QUDEDIS,
Spanish MEC Grant FIS2005-04627, Acciones Integradas and
Consolider QOIT, EU IP SCALA,
the D\'{e}l\'{e}\-gation G\'{e}n\'{e}rale de l'Armement,
the Minist\`ere de la Recherche,
the Agence Nationale de la Recherche,
the European Union, and INTAS.
LCFIO is a member of the Institut Francilien de Recherche sur les Atomes Froids (IFRAF).


\end{document}